\documentclass[journal]{IEEEtran}

\ifCLASSINFOpdf
  \usepackage[pdftex]{graphicx}
\else
\fi

\usepackage{amsmath}
\usepackage{bm}
\usepackage{cite}
\usepackage{amsfonts}
\usepackage{amssymb}

\ifCLASSOPTIONcompsoc
\else
  \usepackage[caption=false,font=footnotesize]{subfig}
\fi

\usepackage{booktabs}
\newcommand{\ra}[1]{\renewcommand{\arraystretch}{#1}}

\begin{document}

	\newcommand{\keyw}[1]{{\bf #1}}
	\newcommand{\reffig}[1]{Fig.\,\ref{#1}} 
	\newcommand{\refeqn}[1]{Eq.\,(\ref{#1})} 
	\newcommand{\refsec}[1]{Section \ref{#1}} 
	\newcommand{\N}{\mathbb{N}} 
	\newcommand{\R}{\mathbb{R}} 

\title{Deep Neural Network Assisted Iterative Reconstruction Method for Low Dose CT}

\author{S. Bazrafkan,
	V. Van Nieuwenhove,
	J. Soons,
	J. De Beenhouwer,
	and J. Sijbers 
	\thanks{S. Bazrafkan, J. De Beenhouwer and J. Sijbers are with imec Visionlab, Department of Physics, University of Antwerp,
		Antwerp, Belgium e-mail: \{shabab.bazrafkan\},\{jan.debeenhouwer\},\{jan.sijbers\}@uantwerpen.be.}
	\thanks{V. Van Nieuwenhove and J. Soons are with Agfa NV, Mortsel, Belgium
		email: \{vincent.vannieuwenhove\},\{joris.soons\}@agfa.com.}}

\maketitle

\begin{abstract}
Low Dose Computed Tomography suffers from a high amount of noise and/or undersampling artefacts in the reconstructed image. In the current article, a Deep Learning technique is exploited as a regularization term for the iterative reconstruction method SIRT. While SIRT minimizes the error in the sinogram space, the proposed regularization model additionally steers intermediate SIRT reconstructions towards the desired output. Extensive evaluations demonstrate the superior outcomes of the proposed method compared to the state of the art techniques. Comparing the forward projection of the reconstructed image with the original signal, shows a higher fidelity to the sinogram space for the current approach amongst other learning based methods.
\end{abstract}


%
\IEEEpeerreviewmaketitle

\section{Introduction}

\IEEEPARstart{C}{omputed} Tomography (CT) is a diagnostic imaging method which operates by acquiring multiple projection images (radiographs) of the object from different angles, after which a 3D image is computed from the set of radiographs. 
CT employs harmful X-ray radiation and there is a continuous strive towards reducing the X-ray dose administered to the patient.  
 
In order to lower the X-ray dose, there are two main approaches. The first one is to reduce the X-ray radiation dose by decreasing the X-ray tube current \cite{mccollough2006ct}. This strategy, however, decreases the Signal to Noise Ratio (SNR) of the projection images and hence also of the reconstructed image.
Another way to decrease the X-ray exposure is to reduce the number of acquired projection images. In other words, the CT device takes images from fewer angles and as a result, a lower amount of radiation is applied during the imaging procedure. This, however, leads to streaking artefacts in the reconstructed image, the severity of which increases with decreasing number of acquired projections. 

In the past decade, the problem of reconstructing images from a small number of projections
has  attracted considerable interest in the field of compressed sensing \cite{Sidky08}. In particular, it was proven
that, if the image is sparse, it can be reconstructed accurately
from a small number of measurements with very high probability, as long as the set of measurements satisfies certain randomization properties \cite{Donoho06}. In many cases, the image itself is not sparse, yet the boundary of the
object is relatively small compared to the total number of pixels.
In such cases, sparsity of the gradient image  can be exploited, by adding a proper regularization term into iterative reconstruction methods \cite{Sidky08}. Similarly, sparsity of the gradient image  \cite{Sidky08}, the grey levels \cite{Batenburg11} or the coefficients in a transform domain \cite{Rantala06} can be exploited to reduce limited data CT artefacts.

In this work, a reconstruction technique based on an iterative method is presented wherein a Machine Learning technique provides prior knowledge that serves as a regularization to the reconstruction.

\subsection{Deep Neural Networks}
In the last few years, Deep Neural Networks (DNN) have played an important role in developing a new generation of Machine Learning techniques known as Deep Learning. These models -if employed in the right place- are able to provide surprisingly high-quality results wherein they already passed the borders of human accuracy on object recognition tasks \cite{he2015delving,ioffe2015batch}. DNNs learn the solution from the training data and generalize this solution to some set of new data they haven't seen before. Typical DNNs consist of several processing units such as Convolutional Layers, Fully Connected Dense Layers, Pooling, and Unpooling layers, which take advantage of techniques such as Batch Normalization \cite{ioffe2015batch} as a regularization step and skipped connections \cite{he2016deep} to keep high frequency information of the input data throughout the network.\\

Currently, we are witnessing a vast number of applications for Neural Networks in several fields of science including low dose CT reconstruction. In \cite{pelt2013fast}, the authors introduce a bank of filters for the FBP method which is learned by a fully connected neural network known as Multi-Layer Perceptron and the weighted sum of several FBP reconstructions is returned as the final result. In \cite{wu2017iterative}, a k-sparse autoencoder \cite{makhzani2015winner} is utilized to learn the priors on the CT data. This model is applied iteratively to perform the reconstruction by applying minimization to the learned manifold alongside with a data fidelity term using a separable quadratic surrogate (SQS) algorithm. In \cite{zhu2018image}, the authors represent an end to end solution wherein the DNN accepts the sinogram space data and generates the reconstructed image at the output. The main problem with this method is the size of the network, which grows exponentially with the input dimension which is the biggest barrier on the practical implementation. The main reason is the two dense fully connected layers wherein every neuron in each layer is connected to every neuron in next and previous layers. The first fully connected layer maps the sinogram to a layer with the size of the output image. The second layer maps this image into an image with the same dimension. These two large layers require a fairly large number of samples to train and still, there are implementation issues considering the required memory. 
In \cite{kang2018deep}, the authors present a framework wherein the denoising is performed in the contourlet space with a network that exploits skipped connection and concatenation layers. The fully convolutional network trained in this work is very large. In fact, in one of the convolutional layers, there are more than 1 million learnable parameters. Such a large network is prone to overfitting if there is not enough representative data available. The main advantage of this work is the mathematical background used in the network design.\\
 
In \cite{kang2017wavelet} the wavelet transform of the reconstructed images is used to train the network in order to perform noise reduction. In other words, the wavelet decomposition is first applied on the reconstructed image and the network repairs the sample, while at the output, the wavelet recomposition is applied. The wavelet transform seems to induce marginal improvements on the final metrics. Methods presented in \cite{chen2017low,chen2017low2,pelt2018improving} train fully convolutional networks to learn noise reduction for FBP reconstruction scenarios. Results from these approaches look promising since the artefact patterns are similar throughout the whole database and the DNNs proved to be efficient in learning these patterns. In the current work, a method similar to these techniques is applied to SIRT reconstruction in a consecutive manner \cite{Gregor2008}.\\
  
Other work includes the technique presented in \cite{yang2018low} wherein a Generative Adversarial Network (GAN) \cite{goodfellow2014generative} is used to learn the distribution of high-quality CT images. Low dose images are then repaired by minimizing a perceptual loss derived from the pre-trained VGG network \cite{simonyan2014very} while the repaired image is forced to have a distribution with minimum Wasserstein distance with the learned distribution. The biggest problem with perceptual loss is its fairness in medical purposes. This is an considerably important issue since every perceptually friendly image might not represent true diagnostic information. In the current work, these loss functions are avoided and Mean Squared Error (MSE) loss has been deployed.\\

The biggest disadvantage of most of the methods described above is the lack of fidelity to the measured sinogram data. This issue is described as follows. The CT device acquires several projections from the object which is also known as the sinogram signal and the construction methods compose an image using any of the aforementioned techniques. If a forward projection is simulated from the reconstructed image, most of these methods do not guarantee that the new sinogram is the same as the sinogram measured by the CT device. In other words, the sinogram of the reconstructed image is different from the original captured signal. This, in fact, is a serious issue which is addressed in this work. The proposed approach reduces the loss in sinogram space by taking advantage of SIRT algorithm which guarantees the fidelity to the measured sinogram and at the same time exploits Deep Learning models to lower the loss in the image space.\\

In the next section, the methodology is described followed by the results and evaluations given in section \ref{sec:results}. Conclusions are presented in the last section.


\section{Methodology}

\subsection{CT Reconstruction}

Let $\bm{x} = (x_j) \in \R^n$ denote the discretised  image of an object, with $n$ the number of pixels.   

In a parallel beam projection geometry, projection data is measured along lines $l_{\theta, t} = \{(x,y) \in \R \times \R: x\cos\theta + y\sin\theta = t\}$, where $\theta \in [0,\pi)$ represents the angle between the line and the $y$-axis and $t$ represents the coordinate along the projection axis.
In practice, a projection is measured at a finite set of projection angles and at a finite set of detector elements, each measuring the integral of the object density along a ray.
Let $\bm{p} = (p_i) \in \R^m$ denote the measured projection data, with $m$ the total number of detector cells times the number of projection angles.

The Radon transform of the object for a finite set of projection directions can be modelled as a linear operator $\bm{W}$, called the \emph{projection operator}, which maps the image $\bm{x}$ to the projection data $\bm{q}$:
\begin{equation}
	\bm{q} := \bm{Wx}.
	\label{eqn:6:qWv}
\end{equation}
In \refeqn{eqn:6:qWv}, $\bm{W}  = (w_{ij})$ is an $m \times n$ matrix where $w_{ij}$ represents the contribution of image pixel $j$ to detector $i$.  The vector $\bm{p}$ is called the \emph{forward projection} or \emph{sinogram} of $\bm{x}$.

The tomographic reconstruction problem can be modelled by the recovery of $\bm{x}$ from a given projection data $\bm{p}$ by solving the following system of equations:
\begin{equation}
	\bm{Wx} = \bm{p}.
	\label{eqn:6:Wvp}
\end{equation}
In the remainder of this work, the \emph{Simultaneous Iterative Reconstruction Technique (SIRT)}, as described in \cite{Gregor2008}, will be used to solve \refeqn{eqn:6:Wvp}.  
SIRT is an iterative algorithm that finds a solution $\bm{\tilde{x}}$ such that the weighted squared projection difference $|| \bm{W\tilde{x}} - \bm{p} ||_{\bm{R}} = (\bm{W\tilde{x}} - \bm{p})^T\bm{R}(\bm{W\tilde{x}} - \bm{p})$ is minimal.  $\bm{R} \in \R^{m\times m}$ is a diagonal matrix that contains the inverse row sums of $\bm{W}$: $r_{ii} = 1/\sum_jw_{ij}$.
The update step of SIRT is given by:
%
\begin{equation}
{\bm x}^{k+1}={\bm x}^k+{\bm C}{\bm A}^T{\bm R}({\bm p}-{\bm W}{\bm x}^k)
\label{eq:SIRTreg}
\end{equation}
where ${\bm x}^k$ is $k$th iteration of the reconstructed image, $\bm C$ is a  diagonal matrices that contain the inverse of the sum of the columns of the system matrix $\bm W$.\\

For the case $\bm{{x}}^{0} = \bm{0}$, the SIRT algorithm is a \emph{linear} algorithm in the sense that a reconstructed image $\bm{\bar{x}} \in \R^n$ is formed by applying a linear transformation to the input vector $\bm{p}$ of projection data.

\subsection{Deep Neural Networks}
Fully convolutional deep neural networks are models that do not contain any dense layers in their architecture. All  layers are convolution, deconvolution, pooling and unpooling which might exploit batch normalization, drop out \cite{dropout}, and/or skipped connections to improve their output quality. The model used in this article is a fully convolutional network wherein just convolutional layers with different dilation sizes and intense skipped connections are used. \\

\subsubsection{Convolutional Layers} 

DNNs are composed of several processing units wherein the convolutional layers play an important role in most of the modern designs. In image processing uses cases, the convolution layers are 3 dimensional: width, height, and channels. The network input is also a 3-dimensional signal in which the samples are stacked in the 4th dimension. While a 4-dimensional kernel maps each layer to the next one, an activation function is applied to the layer output to induce nonlinearity to the model. The following equation describes the mapping of a window in layer $m-1$ to a pixel value in layer $m$.

\begin{equation}
\begin{split}
S^m(x,y,c) = \sigma\Big(\sum_{k=1}^{n_c^{m-1}}\sum_{j=-[n_w/2]}^{[n_w/2]}\sum_{i=-[n_h/2]}^{[n_h/2]}H_{c}^m(i,j,k)\cdot\\
S^{m-1}(x-i,y-j,k)\Big)
\end{split}
\end{equation}

Wherein $S^m(i,j,c)$ is the signal in pixel location $(x,y)$, located in channel $c$ in layer $m$, $H_{c}^m$ is the kernel associated with the channel $c$ of layer $m$. In other words this kernel maps every channel in layer $m-1$ to channel $c$ in layer $m$. $n_h$ and $n_w$ are the width and height of the kernel and $n_{c}^{m-1}$ is number of channels in layer $m-1$. $\sigma$ is the activation function which is also known as the nonlinearity of the layer.\\

\begin{figure}[!t]
\centering
\includegraphics[width=3in]{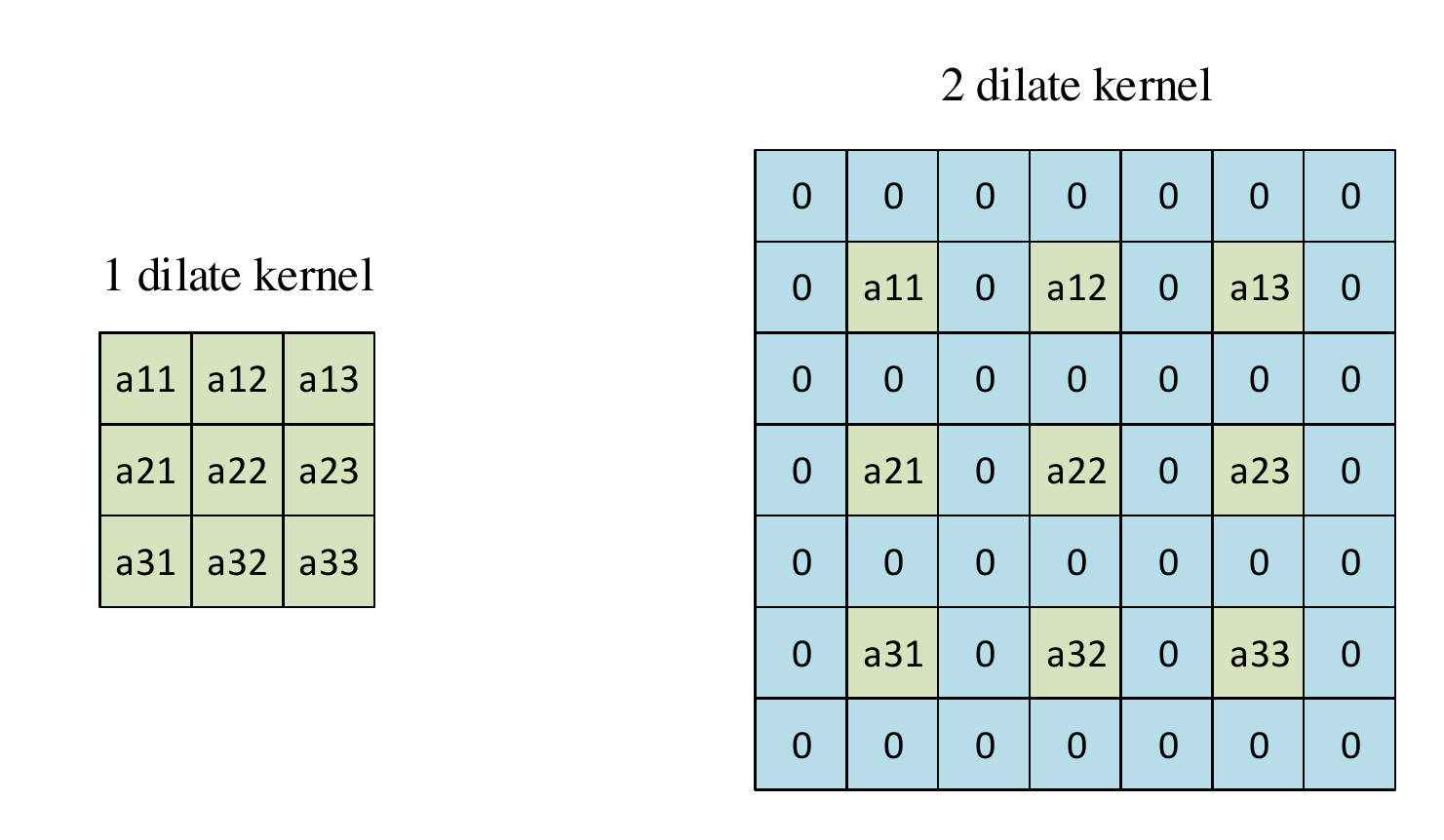}
\caption{$3\times 3$ kernels. Left: 1 dilate. Right: 2 dilate.}
\label{fig:dilate1}
\end{figure}

\begin{figure*}[!t]
\centering
\includegraphics[width=7in]{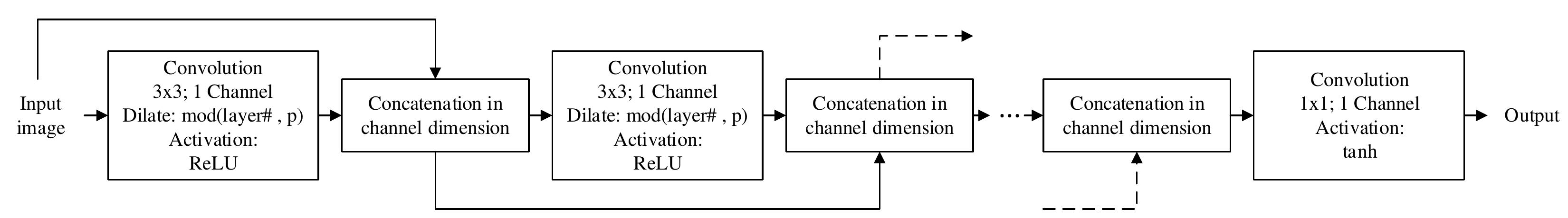}%
\caption{Mixed-Scale Dense Convolutional Network architecture.}
\label{fig:MSD1}
\end{figure*}

\begin{figure*}[!t]
\centering
\includegraphics[width=6.3in]{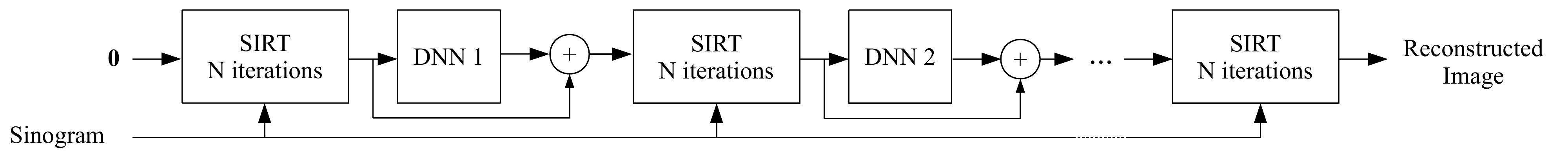}%
\caption{Proposed method for low dose CT reconstruction. DNNs regularize the SIRT output.}
\label{fig:method1}
\end{figure*}

One of the most beneficial properties in convolutional layers is the idea of dilation in the kernel design \cite{dilation}. This quality gives the opportunity for the kernel to increase its field of view while keeping a low number of learnable parameters. The idea is to expand a kernel and fill the void places with zeros as it is illustrated in figure \ref{fig:dilate1}. There are alternative methods such as using larger kernels and/or using pooling operations. A larger kernel means a larger number of parameters which increases the risk of overfitting and the pooling operation induces blurring to the final results. Dilation is a simple and effective approach to increase the receptive field of the kernel without increasing the number of the parameters in the kernel and/or adding pooling layers.\\

\subsubsection{Mixed-Scale Dense (MSD) Convolutional Networks}

The Mixed-Scale Dense (MSD) Convolutional Network is a fully convolutional network which was first introduced in \cite{MSDorig} for image segmentation tasks. Later in \cite{pelt2018improving}, it has been used to remove the low-dose CT reconstruction artefacts of FBP method. MSD structure is shown in figure \ref{fig:MSD1}. This architecture is taking advantage of several dilation scales throughout the model and also because of the single channel convolutions, there are fewer trainable parameters compared to other typical DNNs. In this network, each layer accepts the output of every previous layers concatenated with input image in the channel dimension. The kernel for each layer consists of a $3 \times 3$ convolutional operation. Each kernel has a different dilation value which is specified by the layer number and a value $p$. In the current work $p=10$. All the layers are taking advantage of the well known ReLU nonlinearity \cite{relu}, except the last layer which exploits $tanh$ nonlinearity.\\

The work presented in \cite{pelt2018improving} shows the practicality of MSD in removing streaking artefacts in FBP reconstruction method. The main issue stays to be the fidelity of the method to the sinogram space which is addressed in the following section. 

\subsection{Proposed Method}
As explained in the previous section, the main issue with the current approaches in removing the artefacts in CT imaging is the lack of fidelity to the sinogram space. In other words, there is no guarantee that the sinogram of the reconstructed image matches that of the original sinogram. Iterative methods such as SIRT are designed to decrease weighted squared projection distance in the sinogram space. In fact, these methods minimize the distance between the simulated and measured sinogram. On the other hand, these iterative methods do not guarantee any fidelity in the image space. Depending on the size of the solution space, a reconstructed image can be very different from the scanned object even if the measured sinogram is identical to the simulated sinogram.  At the same time, DNN model requires the image to be as similar as possible to its corresponding ground truth image. In other words, Neural Networks induce the fidelity to the image space. The proposed idea is to use the output of the DNN as the initial point for the SIRT algorithm. In this approach, the DNN steers the SIRT into producing a more realistic output while SIRT ensures that the reconstructed results are entitled to sinogram space fidelity.
In order to accomplish this, the DNN is utilized as a regularization unit for SIRT. The following equation shows the update stage for SIRT including the regularization term.
\begin{equation}
{\bm x}^{k+1}={\bm x}^k+{\bm C}{\bm A}^T{\bm R}({\bm p}-{\bm W}{\bm x}^k)+\text{REG.}
\label{eq:SIRTreg}
\end{equation}

The idea is to provide a DNN which regularizes the term `REG' in a way that:
\begin{equation}
{\bm x}^{k+1}=\text{GT.}\quad,
\label{eq:XGT}
\end{equation}
wherein GT is the ground truth of the reconstructed image. It means that the regularization term forces SIRT to provide a perfect reconstruction in the image space. From equations (\ref{eq:SIRTreg}) and (\ref{eq:XGT}) it is concluded that:

\begin{equation}
\text{REG} = \text{GT} - \big({\bm x}^k+{\bm C}{\bm W}^T{\bm R}(\bm p-{\bm W}{\bm x}^k)\big).
\label{eq:REG}
\end{equation}

Equation (\ref{eq:REG}) implied that the regularization term should provide the residual value between the reconstructed image and the ground truth. The proposed method is illustrated in figure \ref{fig:method1}. Several DNNs will be trained with the reconstruction image as the input and residual value as target. At the inference step, the trained DNN provides the residual value which is used to generate a new initialization point for SIRT. The regularization is applied once in every $N$ SIRT iterations.

\subsection{Data Base}
{\bf CPTAC-PDA}: National Cancer Institute’s Clinical Proteomic Tumor Analysis Consortium Pancreatic Ductal Adenocarcinoma (CPTAC-PDA)\footnote{https://wiki.cancerimagingarchive.net/display/Public/CPTAC-PDA} is a publicly available database containing  45786 Pancreas images from CPTAC phase 3 patients. It consists of 45 radiology and 77 pathology subjects. This database contains several modalities including CT, Computed Radiography (CR) and MRI samples. Images are from different sizes but in the current work, they were resized to $128\times 128$. Using multiple modalities in the training stage increases the generality of the solution induced by the different properties of various imaging techniques.\\

{\bf Visible Human Project CT Datasets}: Visible Human Project CT Datasets\footnote{https://mri.radiology.uiowa.edu/visible\_human\_datasets.html} contains 2989 images from 10 CT imaging cases. This dataset is publicly available. Images are $512\times 512$ while in the current study they were all resized to $128\times 128$. This database contains CT images of the ankle, head, hip, knee, pelvis, and shoulder from both male and female subjects. The male shoulder samples (461 images) were isolated from all training data to be used as the test set.\\

The low dose scenario is simulated by taking a limited number of projections from every image in the database. A parallel beam geometry \cite{astra3} with 20 equidistant projections between 0 and 180 degrees has been utilized to produce the low dose sinogram. The ASTRA Toolbox\footnote{https://www.astra-toolbox.com/} \cite{astra1,astra2} provides the required tools in order to simulate the X-ray projections.
In this study, the male shoulder samples from Visible Human Project CT Datasets  (461 images) are used as test set and the rest of the data is employed in the training procedure. 80\% of the training dataset is used for network training and the remaining 20\% for validation.

\subsection{Training}
\begin{figure}[!t]
\centering
\includegraphics[width=3.5in]{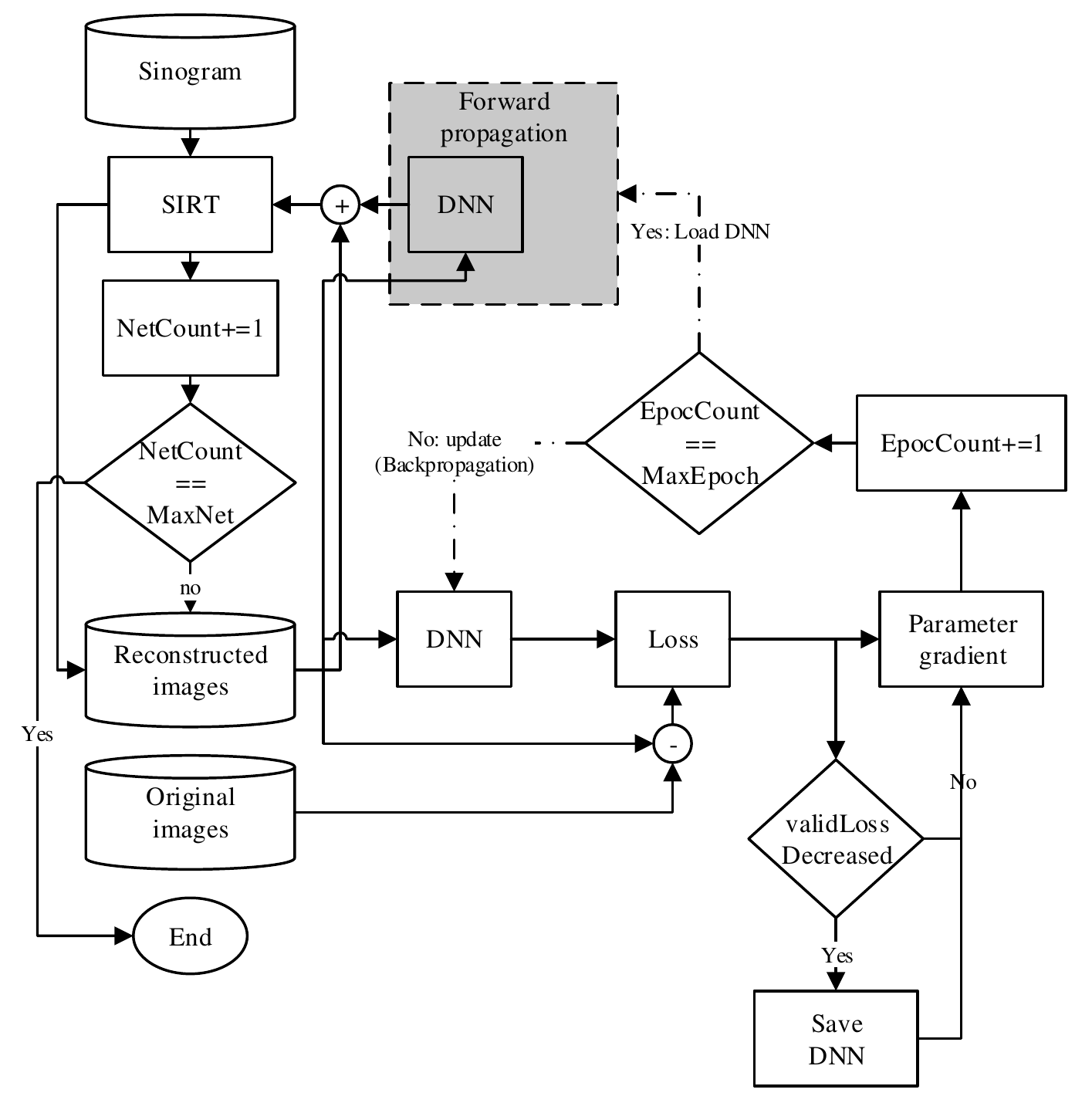}
\caption{Training procedure for the proposed framework.}
\label{fig:training1}
\end{figure}

In order to obtain the framework shown in figure \ref{fig:method1}, several DNNs are trained consequently. In this work, the regularization is applied 10 times so there are 10 different networks trained in figure \ref{fig:method1}. An MSD convolutional architecture used for all networks. The network consists of 51 layers with $p=10$. The number of SIRT iterations before each regularization step is $N=10$. The training procedure is shown in figure \ref{fig:training1}, with $MaxNet=10$ and $MaxEpoches=100$. $MaxNet$ is the number of the networks and $MaxEpoches$ is the maximum number of epochs each network is trained. The first DNNs parameters were initialized uniformly in the range $[-0.25,0.25]$ and the further networks were initialized from the model saved for the previous step. This technique which knows as transfer learning has been widely used in various applications and it is certainly effective in the current problem wherein the artefacts induced by SIRT in different steps are quite similar.\\

The Mean Squared Error is used as the loss function for training which is given by:

\begin{equation}
Loss = \frac{1}{B_s H W}\sum_{k=1}^{B_s}\sum_{j=1}^{H}\sum_{i=1}^{W}\big(O(i,j,k)-t(i,j,k)\big)^2\quad,
\label{eq:MSE}
\end{equation}
where $W$, $H$, and $B_s$ are the width, height and the batch size of the input signal, respectively. A batch size equal to 10 has been used in this work. An ADAM optimizer \cite{adam} have been utilized to update the parameters with learning rate, $\beta_1$, $\beta_2$ and $\epsilon$ equal to $0.0001$, $0.9$, $0.999$, and $10^{-8}$ respectively. The MXNET 1.3.0 \cite{mxnet} \footnote{https://mxnet.apache.org/} framework have been used to train the network and the ASTRA Toolbox \cite{astra1,astra2} was used to perform the SIRT step. The training was accomplished on one TESLA V100 \cite{teslav100} GPU of a DGX Station \cite{dgxstation}.\\

\begin{figure*}[!t]
\centering
\subfloat[Training Loss]{\includegraphics[width=3.5in]{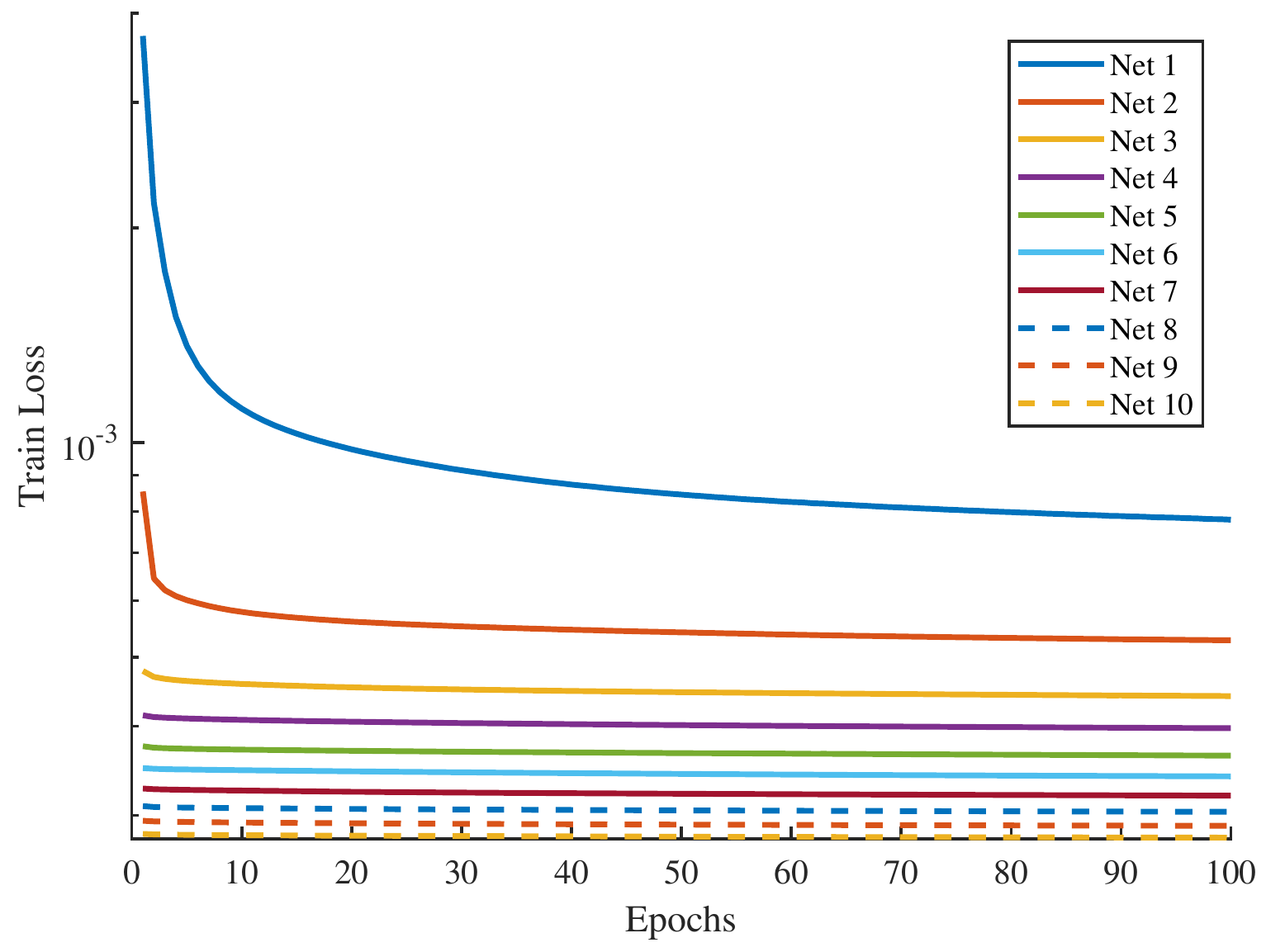}%
\label{fig:TrainLoss}}
\hfil
\subfloat[Validation Loss]{\includegraphics[width=3.5in]{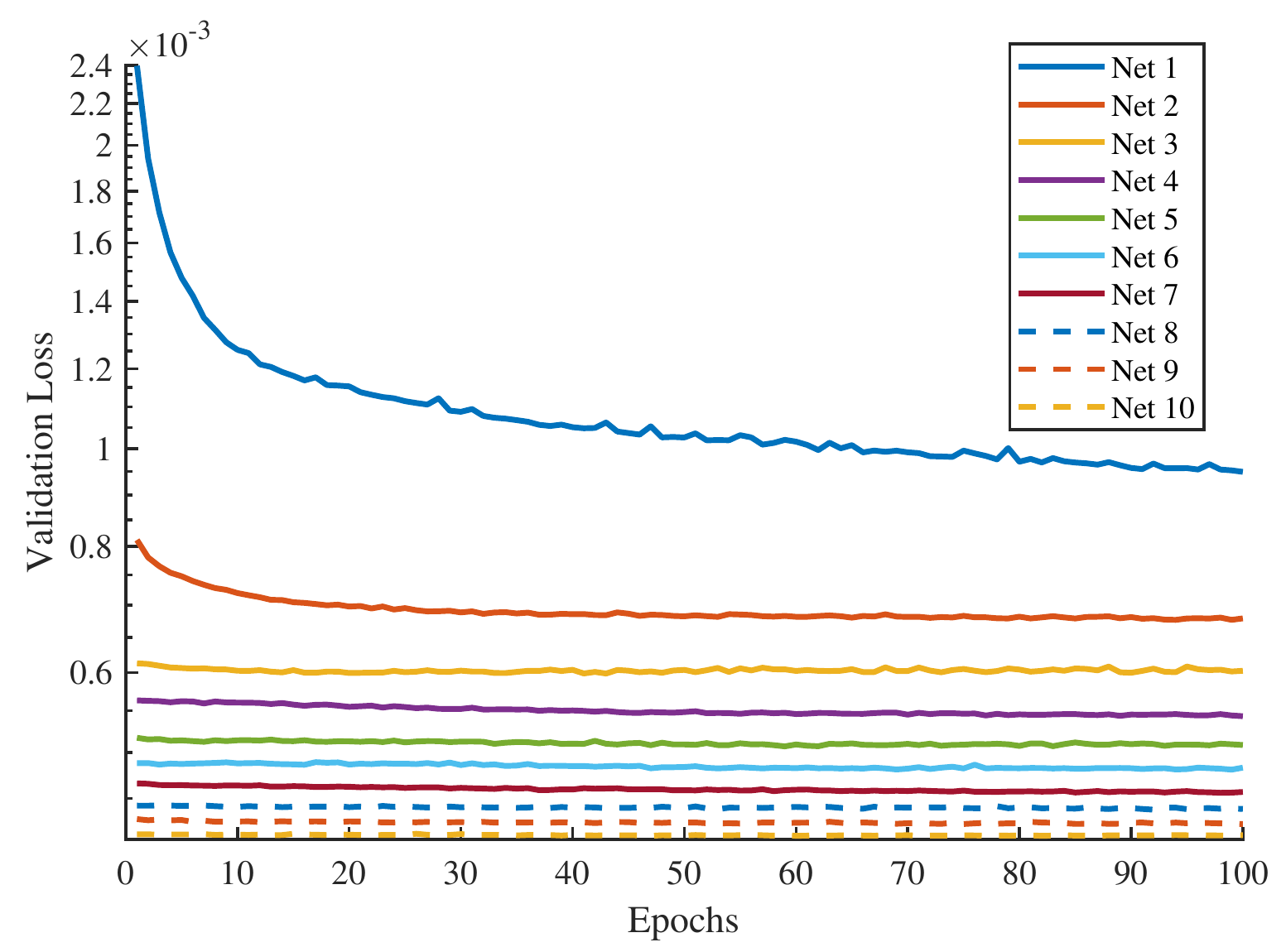}%
\label{fig:ValidationLoss}}
\caption{Training and validation losses for 10 networks.}
\label{fig:Losses}
\end{figure*}

Figures \ref{fig:TrainLoss} and \ref{fig:ValidationLoss} illustrates the train loss and validation loss for each of ten trained networks. As it is shown, the network losses decrease after each SIRT block. This is the sign of cooperative behaviour of SIRT and DNN, while SIRT minimizes the error in sinogram space it also decreases the image space loss thanks to the regularization term provided by the DNN. The same improvement is visible in the validation loss which declares the generalization of the method.  

\section{Results}
\label{sec:results}

 In this section, the proposed method is evaluated and compared to the state of the art methods in the literature. The term SIRT+DNN is used to represent the current method. The other techniques used for comparison are as follows:
 
\begin{enumerate}
\item Model-based methods:\\
Filtered Back Projection (FBP), Simultaneous Iterative Reconstruction Technique (SIRT) \cite{Gregor2008}, Conjugate Gradient Least Squares (CGLS) \cite{Paige82}, and  Total Variation with adaptive step size (TV adaptive) \cite{Yokota17}.\\

\item Learning based methods:
\begin{enumerate}
\item {\bf AUTOMAP} \cite{zhu2018image}: This is the state of the art implementation of an end to end design for image reconstruction, wherein a single DNN is trained to perform this job. It accepts the sensor signal and returns the reconstructed image. In this work, the low dose sinogram signal is used as input and the original image as the target image to provide the loss function. The AUTOMAP network is trained on the exact same data as the SIRT+DNN.  The Mean Squared Error is used as the loss function with ADAM optimizer updating the network parameters. The learning rate, $\beta_1$, $\beta_2$, and $\epsilon$ are set to $0.0001$, $0.9$, $0.999$, and $10^{-8}$, respectively. The main disadvantage with this method is the fact that it is a fully data-driven method which does not take advantage of the image acquisition and geometry properties.\\ 

\item {\bf FBP+DNN}: This method utilizes a DNN to learn the artefacts induced by the FBP method. This technique has been  widely used in the literature and in the current evaluations, the framework presented in \cite{pelt2018improving} has been employed. In order to provide a fair comparison, the FBP+DNN model has been trained on the same MSD network as SIRT+DNN. The same database is used in the training procedure. The Mean Squared Error is used as the loss function as declared in \cite{pelt2018improving}. The ADAM optimizer is utilized to update the parameters with learning rate, $\beta_1$, $\beta_2$, and $\epsilon$ equal to $0.0001$, $0.9$, $0.999$, and $10^{-8}$, respectively.\\

\item {\bf Neural Network Filtered Back Projection NNFBP (16,32,64)} \cite{pelt2013fast} In this approach, a fully connected neural network is trained to find a set of best filters for the FBP method. The numbers in the parenthesis declared the number of hidden units deployed in the network. In the observations done in \cite{pelt2013fast} it has been shown that the networks with 16, 32 and 64 hidden layers return the results with highest accuracies, therefore, these three setups have been used in the current evaluation section. Since the training is performed on the pixel level, the test set has been used for training \footnote{http://dmpelt.github.io/pynnfbp/}. This gives the opportunity to compare the proposed method with the best version of NNFBP on the current data.\\

\end{enumerate} 
\end{enumerate}

\begin{table*}\centering
\ra{1.3}
\begin{tabular}{@{}rcccccccc@{}}\toprule
& \multicolumn{2}{c}{PSNR} & \phantom{abc}& \multicolumn{2}{c}{MSE} &
\phantom{abc} & \multicolumn{2}{c}{SSIM}\\
\cmidrule{2-3} \cmidrule{5-6} \cmidrule{8-9}
& $\mu$ & $\sigma$ && $\mu$ & $\sigma^2$ && $\mu$ & $\sigma$ \\ \midrule
{\bf Model Based Methods}\\
FBP & 23.7 & 1.2  && 4.4e-3 & 1.0e-3  && 0.6015 & 2.5e-2 \\
CGLS & 29.3 & 1.0  && 1.2e-3 & 2.6e-4  && 0.8310 & 2.2e-2 \\
SIRT & 28.8 & 1.1  && 1.3e-3 & 3.0e-4  && 0.8148 & 2.3e-2 \\
TV Adaptive & 29.7 & 1.2  && 1.1e-3 & 2.6e-4  && 0.8532 & 2.1e-2 \\
{\bf Learning Based Methods}\\
FBP+DNN & 31.0& 1.1 && 8.1e-4& 1.8e-4&& 0.9020& 1.5e-2\\
NNFBP16 & 25.1& 1.1 && 3.2e-3& 8.6e-4&& 0.8050& 2.1e-2\\
NNFBP32 & 27.8& 1.5 && 1.8e-3& 7.4e-4&& 0.8350& 1.9e-2\\
NNFBP64 & 29.9& 0.9 && 1.0e-3& 2.3e-4&& 0.8542& 1.7e-2\\
AUTOMAP & 28.2& {\bf 0.8} && 1.5e-3& 2.9e-4&& 0.8549& 1.6e-2\\
SIRT+DNN & {\bf 37.2} & 1.3  && {\bf 1.9e-4} & {\bf 5.1e-5}  && {\bf 0.9805} & \bf{4.4e-3} \\
\midrule
\label{tab:ImageSpace}
\end{tabular}
\caption{Evaluations in the image space on PSNR, MSE and SSIM}
\end{table*} 

The measurements used for evaluations are as follows:
\begin{enumerate}
\item {\bf Peak Signal to Noise Ratio (PSNR)}:  Is the ratio between the maximum power of the signal to the power of the noise. This measure is widely used in image comparison for reconstruction use cases. The higher value indicates better reconstruction quality.\\

\item {\bf Mean Squared Error (MSE)}: Represents the power of the noise in the reconstructed image. The lower value of MSE corresponds to higher quality reconstruction. Both MSE and PSNR are pixel level measures. In other words, these measures calculate the difference between two images in the pixel level grayscale values.\\ 

\item {\bf Structural Similarity Index (SSIM)}: is a quality measurement presented in \cite{SSIM} wherein two images are compared based on their structural information and not solely the pixel value. The index range is between zero and one which zero indicates no similarities and one is a perfect structural match between the reconstructed image and ground truth. 
\end{enumerate}

\subsection{Image Space Evaluations}

\begin{figure*}[!t]
\centering
\includegraphics[width=5in]{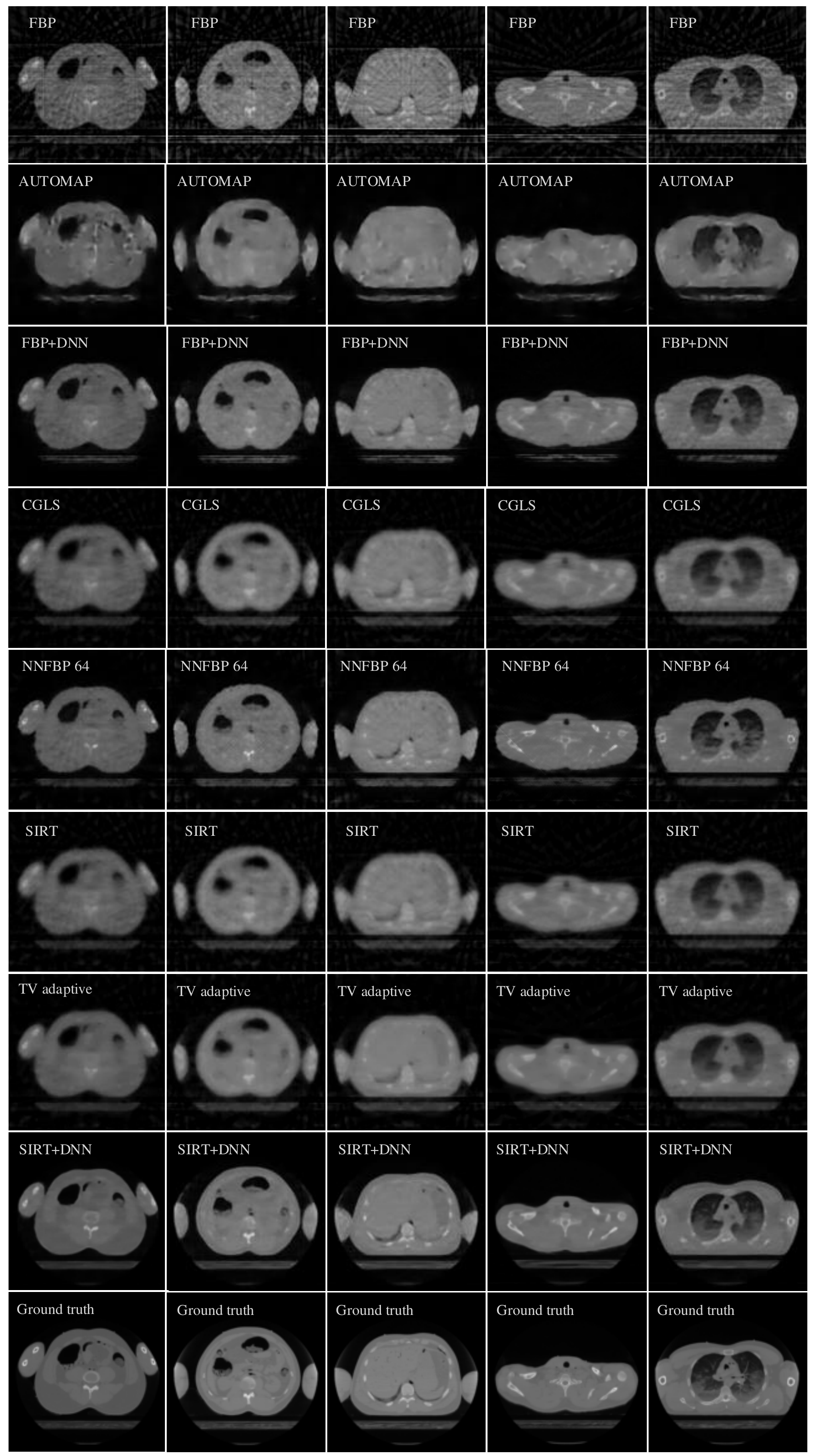}%
\caption{Reconstruction examples taken from the test set alongside their corresponding ground truth}
\label{fig:results1}
\end{figure*} 

\begin{figure*}[!t]
\centering
\includegraphics[width=4.7in]{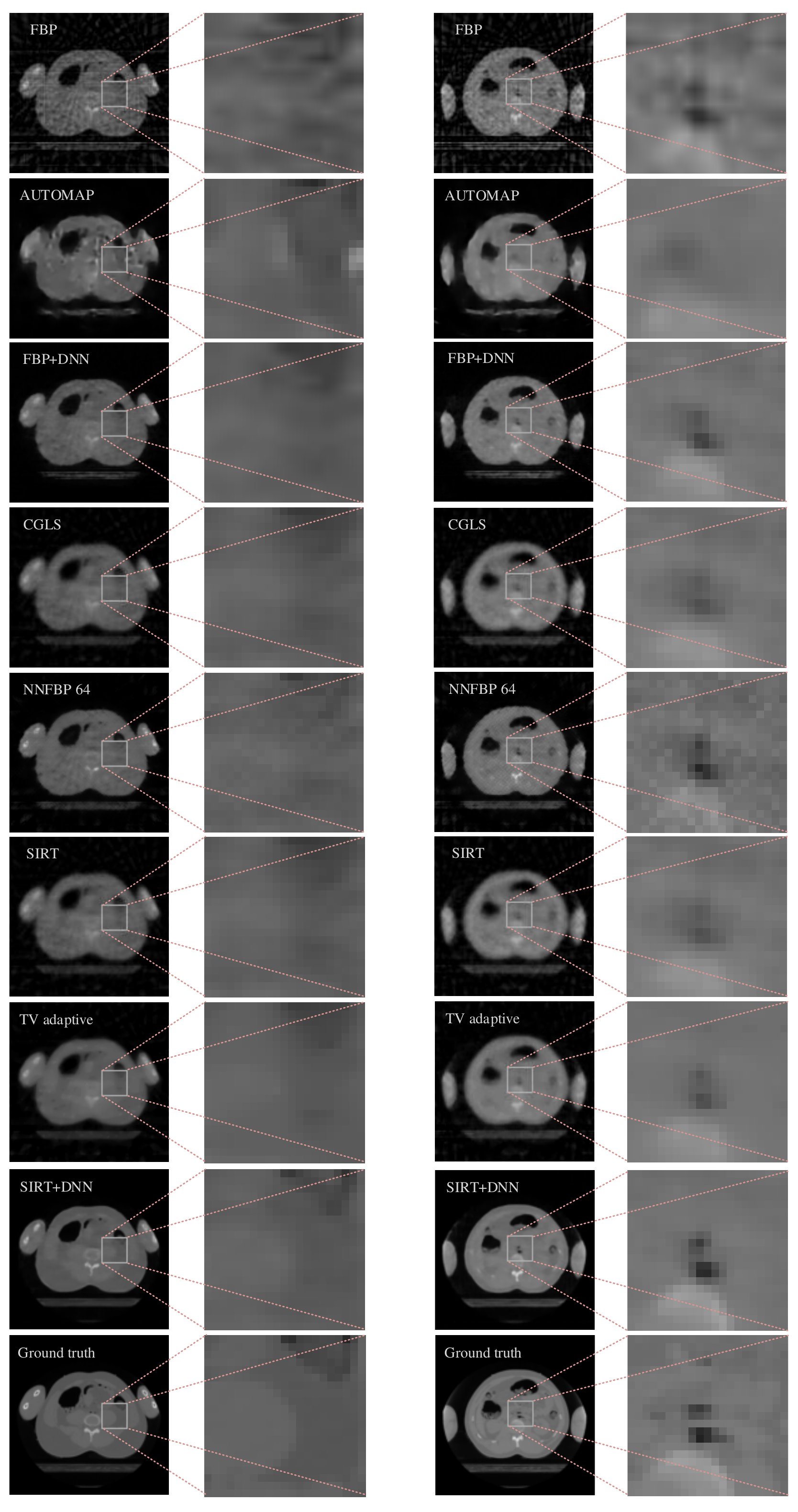}%
\caption{Left: soft tissue reconstruction examples. Right: sharp edges examples.}
\label{fig:results2}
\end{figure*} 

Table I shows the comparisons between the proposed method and the state of the art methods in the literature. High PSNR and low MSE declare a higher accuracy in returning pixel level information for the proposed scheme.  And the high SSIM value shows the consistency of SIRT+DNN in keeping the structural information even in very low CT doses. The learning based methods return a higher SSIM value especially when they are used as an auxiliary step to remove artefacts from the reconstruction image as in FBP+DNN and SIRT+DNN techniques. This is while SIRT+DNN delivers the highest PSNR and lowest MSE compared to other methods. As shown in figure \ref{fig:method1} each SIRT step reduces the loss in the sinogram space which in practice induces streaking artefacts to the image space and each DNN steps removes these artefacts without considering the consistency in the sinogram space. The mixture of these two steps consecutively provides a strong tool in returning a high-quality reconstruction in both pixel level and structural information. Figure \ref{fig:results1} illustrates the reconstruction results for the provided geometry (20 projections parallel beam) in different methods. \\

The AUTOMAP method does not introduce any streaking but the provided images are suffering from a tangled artefact. In other words, the details are mostly twisted together which results in a matte reconstruction. The main reason for this behavior is that AUTOMAP does not include any geometry information in the reconstruction procedure and also the low dose scenario increases the uncertainty of the solution. All other methods suffer from severe streaking artefacts and even in FBP+DNN scheme, the network is not able to remove all the artefacts. In other cases such as CGLS, SIRT, and TVadaptive, a certain amount of blurring is also introduced to the image. Especially in the TVadaptive method, details are merged into a single block due to the total variation term of the loss function. \\

Figure \ref{fig:results2} illustrates zoom up images which give more elaborate insight into the presented technique. In the left column, a soft tissue block is illustrated. Most of the methods fail to reconstruct the correct low contrast property of the soft tissue. AUTOMAP clearly fails to produce even the slightest structural information. Other methods such as SIRT, FBP, CGLS, and NNFBP return a blurry image which suffers from the lack of any recognizable edges. The next best result is from the TVadaptive method which produces sharper edges when the contrast is large enough (air/object or bone/soft tissue transitions) but yields severely blurred results in the soft tissue regions. The SIRT+DNN method returns a better output for the soft tissue. There is a similar situation for the right column images where correspond to the reconstruction of sharp edges. The FBP and NNFBP methods induce a strong level of noise into the image which is very difficult to remove even with a DNN. This is shown in the FBP+DNN image in which the black region is closed due to the filtering applied in the DNN step. Again, the proposed SIRT+DNN method produces the best reconstruction compared to the other methods.

\subsection{Sinogram Space Evaluations}
\begin{figure}[!t]
\centering
\includegraphics[width=3.5in]{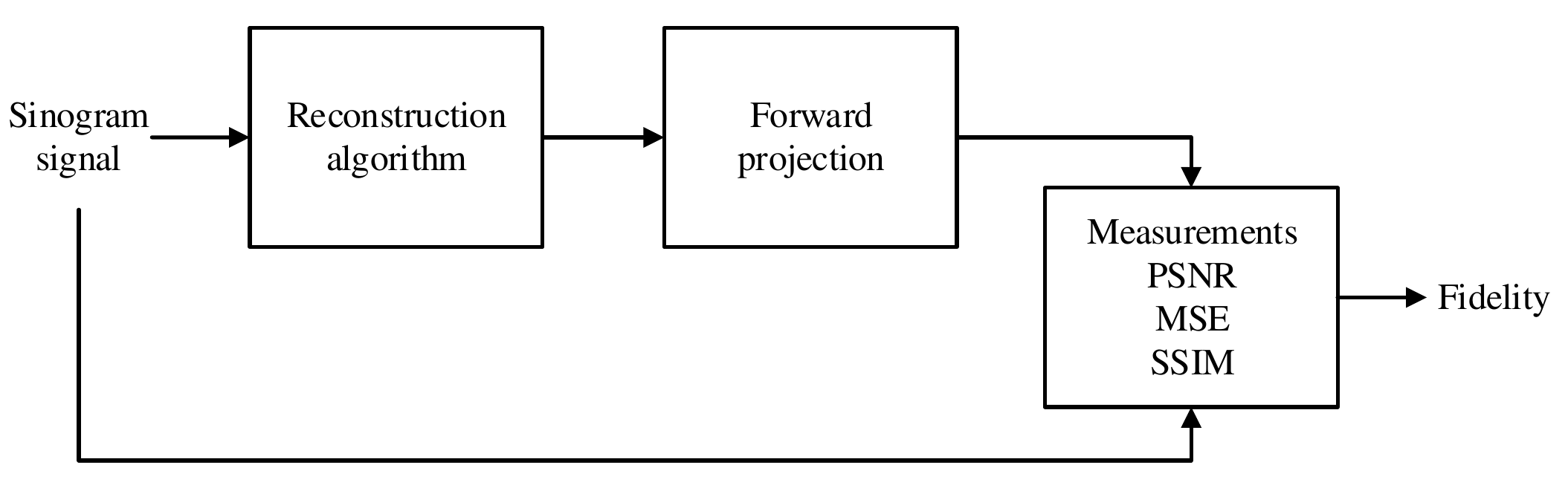}%
\caption{Processing blocks for measuring sinogram fidelity.}
\label{fig:fidelityblock}
\end{figure} 

\begin{table*}\centering
\ra{1.3}
\begin{tabular}{@{}rcccccccc@{}}\toprule
& \multicolumn{2}{c}{PSNR} & \phantom{abc}& \multicolumn{2}{c}{MSE} &
\phantom{abc} & \multicolumn{2}{c}{SSIM}\\
\cmidrule{2-3} \cmidrule{5-6} \cmidrule{8-9}
& $\mu$ & $\sigma$ && $\mu$ & $\sigma$ && $\mu$ & $\sigma$ \\ \midrule
{\bf Model Based Methods}\\
FBP & 39.1 & 1.3  && 1.3e-4 & 3.1e-5  && 0.9686 & 5.7e-3 \\
CGLS & {\bf 101.9} & {\bf 1.0}  && {\bf 6.6e-11} & {\bf 1.5e-11}  && {\bf 1.0000} & {\bf 3.9e-9} \\
SIRT & 92.2 & 2.1  && 7.3e-10 & 1.1e-9  && {\bf 1.0000} & 3.1e-7 \\
TV Adaptive & 61.8 & 1.2  && 6.9e-7 & 1.7e-7  && 0.9999 & 3.4e-5 \\
{\bf Learning Based Methods}\\
FBP+DNN & 47.7& 1.8 && 1.8e-5& 7.3e-6&& 0.9918& 3.0e-3\\
NNFBP16 & 28.2& 1.6 && 1.6e-3& 6.2e-4&& 0.9479& 1.1e-2\\
NNFBP32 & 28.8& 1.6 && 1.4e-3& 5.6e-4&& 0.9508& 2.6e-2\\
NNFBP64 & 28.5& 1.7 && 1.5e-3& 6.4e-4&& 0.9535& 1.4e-2\\
AUTOMAP & 45.4& 1.2 && 3.0e-5& 8.2e-6&& 0.9923& 1.6e-3\\
SIRT+DNN & {\bf 68.7} & {\bf 0.9}  && {\bf 1.4e-7} & {\bf 2.8e-8}  && {\bf 1.0000} & \bf{5.6e-6} \\
\midrule
\end{tabular}
\label{tab:SinogramSpace}
\caption{Evaluations in the sinogram space on PSNR, MSE and SSIM}
\end{table*}

As described earlier most of the learning based reconstruction methods suffer from the lack of fidelity to the sinogram space. In other words, the forward projection of the reconstructed image differs from the original measurements taken from the sensors. In order to overcome this drawback, the proposed method takes advantage of SIRT which decreases the loss in sinogram space. And the DNN is utilized as a regularization term which introduces image space information into the model.\\

The PSNR, MSE, and SSIM measurements are calculated from the pipeline shown in figure \ref{fig:fidelityblock}. This is done for all the methods and the results are shown in table II. The model-based techniques such as SIRT and CGLS are designed to explicitly keep the reconstruction sinogram as close as possible to the measurements. TV technique imposes corrections in the image space which reduces its fidelity to the sinogram space.  The FBP method does not couple back to the sinogram which it is why it returns the worst fidelity among the model-based methods. \\

In the learning based methods, the NNFBP returns the worst values which indicate that the designed filters do not require sinogram fidelity. AUTOMAP and FBP+DNN give the next best results. In fact, getting a higher value for FPB+DNN compared to the original FBP shows that the DNN is pushing the results towards a higher sinogram fidelity. This is happening while the sinogram loss term is not included in the DNN objectives. \\

The proposed SIRT+DNN method produces the best results in sinogram space among the learning based methods. It is also worthwhile to mention that while DNN does not improve the sinogram fidelity compared to SIRT but it has a significant impact on the image space measurements.

\subsection{Sinogram Noise Evaluation}

\begin{figure}[!t]
\centering
\includegraphics[width=3.5in]{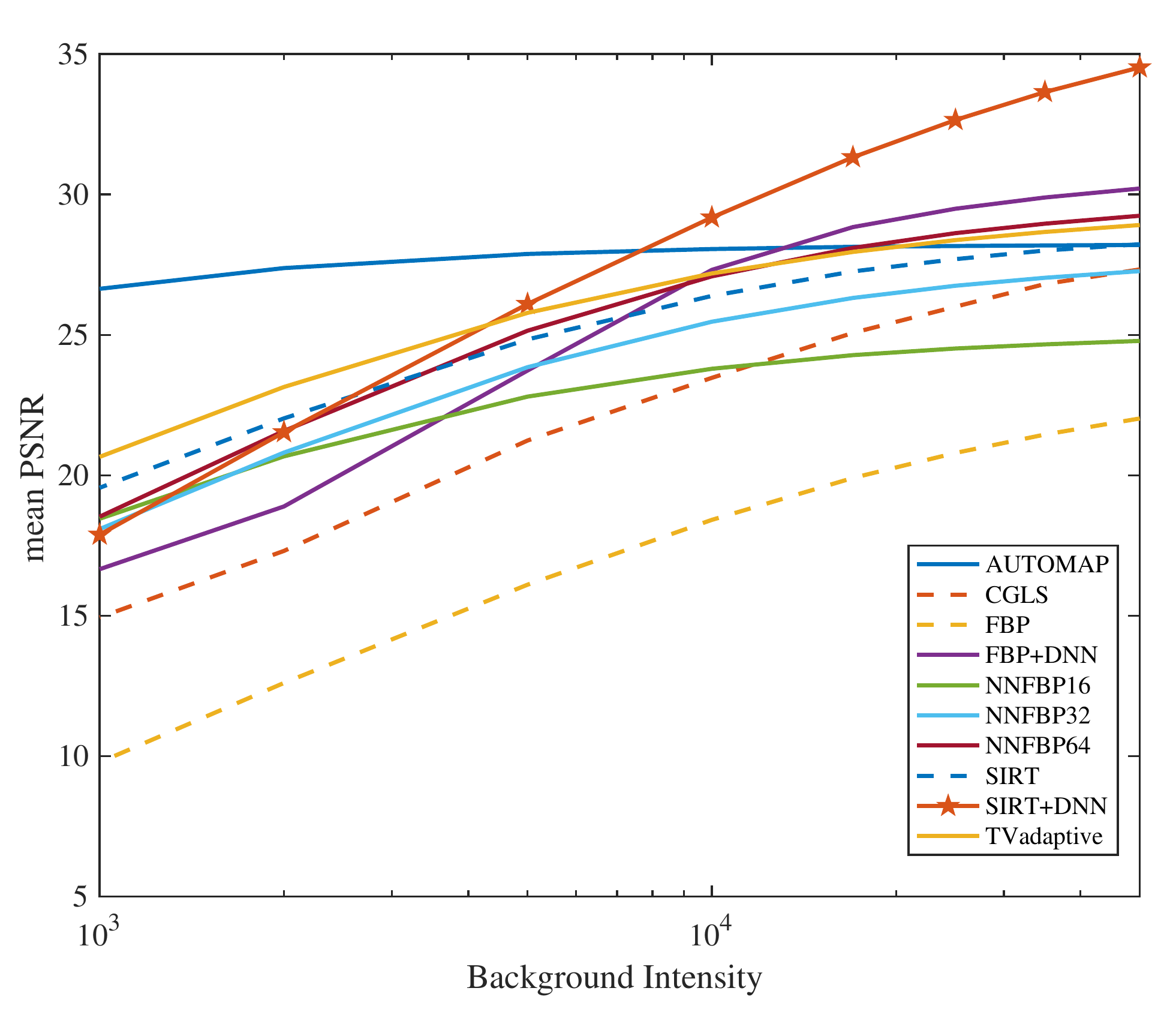}%
\caption{Mean PSNR wrt the background intensity for different methods.}
\label{fig:noisepsnr}
\end{figure} 

\begin{figure}[!t]
\centering
\includegraphics[width=3.5in]{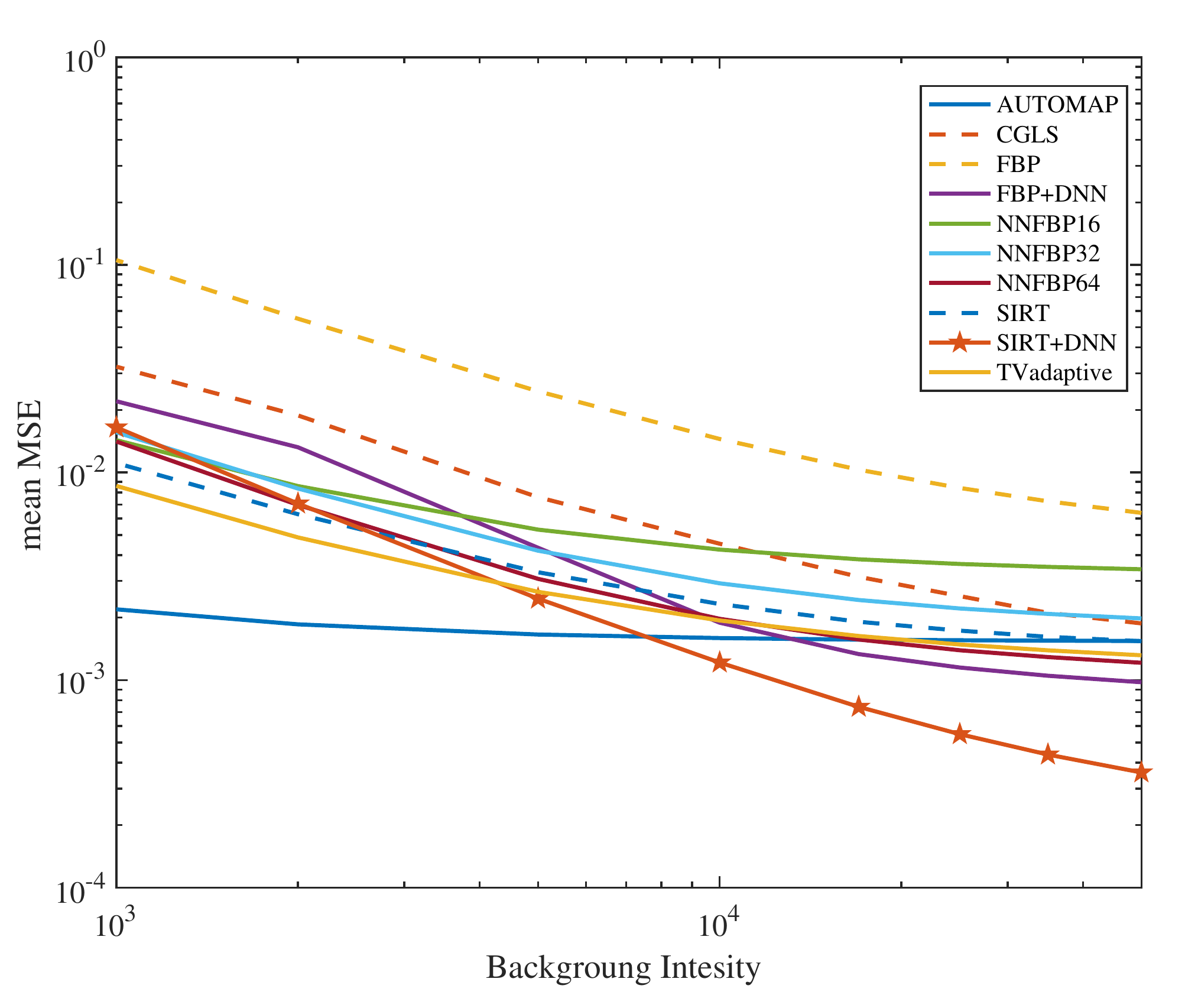}%
\caption{Mean MSE wrt the background intensity for different methods.}
\label{fig:noisemse}
\end{figure}

\begin{figure}[!t]
\centering
\includegraphics[width=3.5in]{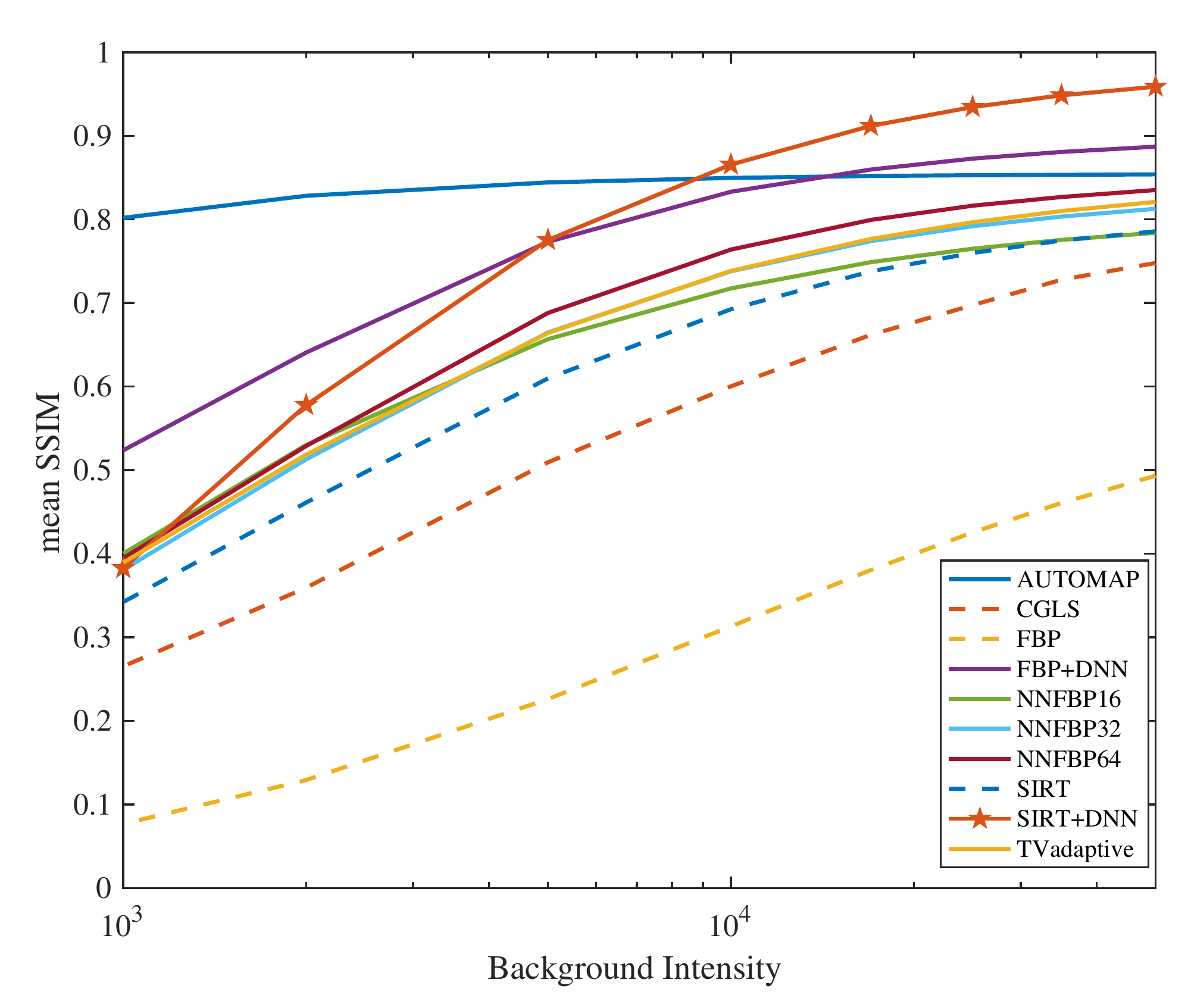}%
\caption{Mean SSIM wrt the background intensity for different methods.}
\label{fig:noisessim}
\end{figure}

To investigate the performance of the proposed reconstruction method in terms of the noise in the sinogram, projection
data of each phantom image was generated to which Poisson noise was applied. The intensity of this noise is defined by the incident beam intensity, $I_0$ (further referred to as the background intensity), i.e. the photon count in the incident X-ray beam. Reconstructions were performed using different values for $I_0$.\\

The PSNR, MSE, and SSIM of the reconstructed images as a function of the noise level in the projection images (by varying the background intensity) for several methods are illustrated in figures \ref{fig:noisepsnr} to \ref{fig:noisessim}, respectively. At very low background intensities (high noise power), the learning based methods are able to keep the structural information better than model-based methods. Considering pixel level information, SIRT and TVadaptive give higher quality results compared to FBP and CGLS in high noise power. AUTOMAP gives the most robust results giving different noise levels considering that it is a fully learning based method and no model information is utilized in developing this model. The proposed SIRT+DNN method lies within NNFBP methods in low background intensity but it has the highest slope in improving the output with respect to the noise level. In other words for background intensities higher than 10000, the proposed method returns superior results compared to the other techniques. It is worthwhile to mention that no amount of noise was added to the input samples in the training stage. The results for the learning-based methods will improve by providing broader range of variations in the training set including noisy samples.

\subsection{Intermediate Results}
\begin{figure*}[!t]
\centering
\includegraphics[width=7in]{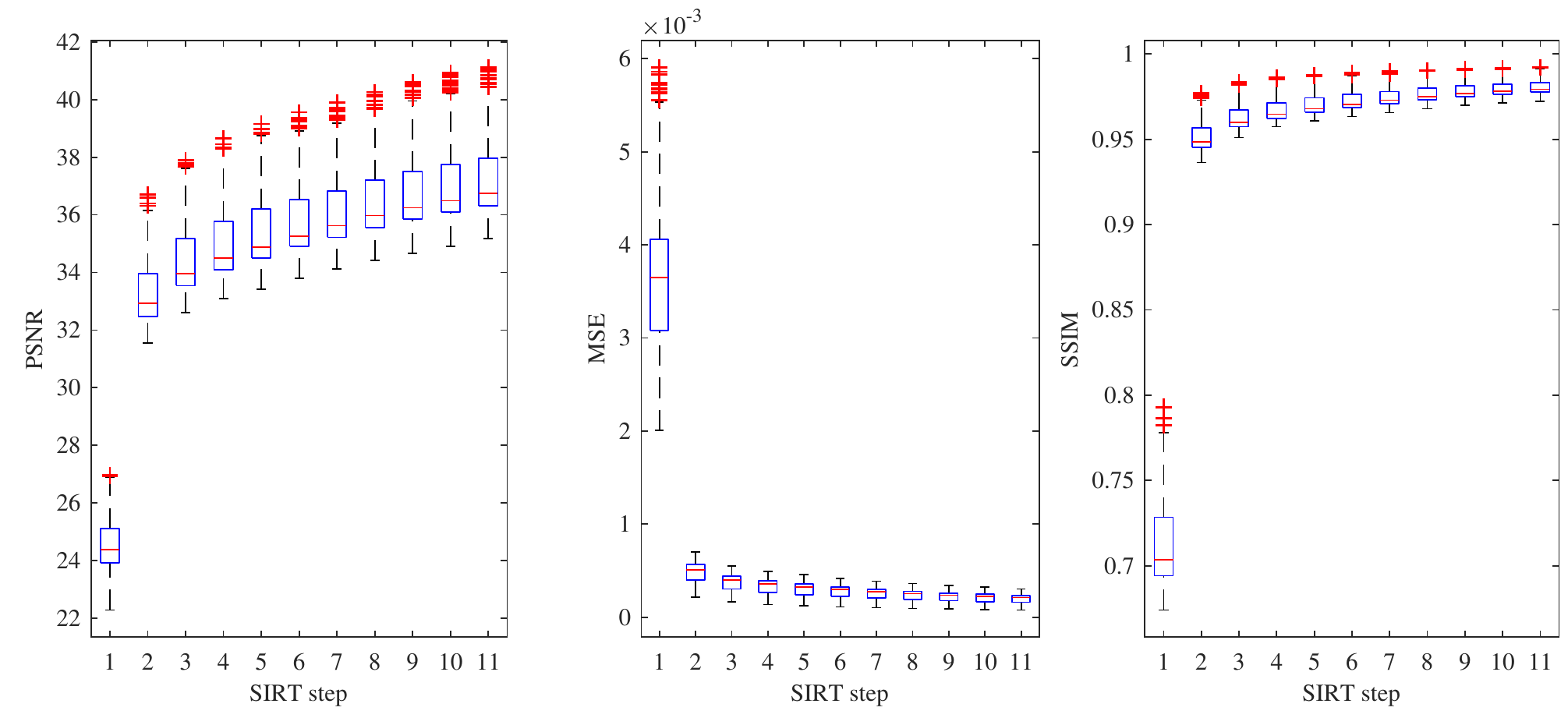}%
\caption{Test statistics after each SIRT step for PSNR, MSE and SSIM measurements.}
\label{fig:intermediateaftersirt}
\end{figure*}

\begin{figure*}[!t]
\centering
\includegraphics[width=7in]{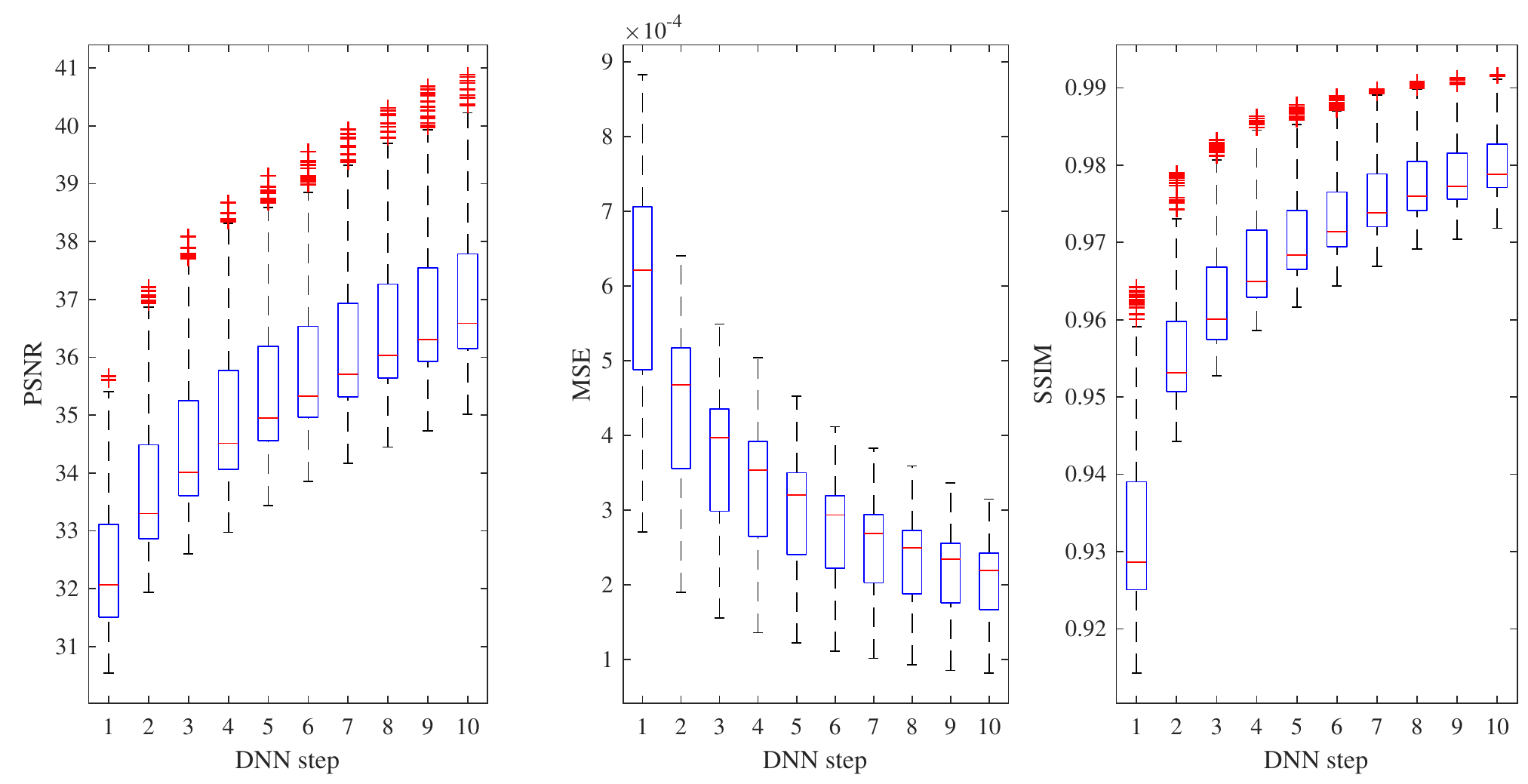}%
\caption{Test statistics after each DNN step for PSNR, MSE and SSIM measurements.}
\label{fig:intermediateafterdnn}
\end{figure*}

As shown in figure \ref{fig:method1} the proposed method consists of several SIRT and DNN blocks placed consecutively. In this section, the results after each step are investigated. The considered blocks are divided into two observations for SIRT  and DNN individually. In the current simulations, ten DNNs have been trained, one after each SIRT step and at the end, a final SIRT was applied to the network output. Therefore there are ten DNN and eleven SIRT blocks in total. The output of each of these blocks are calculated for the test set and PSNR, MSE and SSIM measures are plotted in figures \ref{fig:intermediateaftersirt} and \ref{fig:intermediateafterdnn} after each SIRT and DNN block respectively. Both pixel value and structural features improve after each SIRT and DNN step. This indicates the cooperative behavior of SIRT and DNN. In other words, these figures show that improvements induced by DNN are in the same direction of optimizing the loss in the sinogram space. It is also shown that the early stages of the model play an important role in the whole workflow. The low-quality results after the first SIRT block are highly impacted by the first DNN block, while as going forward in the processing steps, the improvements get more and more marginal. It is also worthwhile to mention that these improvements generate a more detailed reconstruction which yields the proposed SIRT+DNN method to stand out amongst other reconstruction methods.

\section{Conclusion}
In this article, a technique for low dose CT reconstruction has been proposed wherein a Deep Neural Network is utilized as a regularization term for a classical iterative reconstruction algorithm known as SIRT. Ten Mixed-Scale Dense Convolutional Deep Neural Networks have been employed consecutively after ten SIRT blocks. The first network has been initialized randomly and further networks take advantage of the transfer learning technique wherein each network is initialized as the best network from the previous step.\\
 
In the results section, the proposed method is compared to state of the art techniques in CT image reconstruction where it shows a superior improvement in PSNR, MSE and SSIM measurements compared to other methods.\\
 
Another problem tackled in the proposed technique is the fidelity to the sinogram space. Most of the learning based methods act in the image space which is blind to the sinogram space and the reconstructed image after forward projection differs from the originally measured sinogram. By using the power of SIRT in decreasing the loss in sinogram space and the DNN optimizing the model in image space, SIRT+DNN returns the best measures in sinogram fidely amongst other learning-based methods alongside the superior results in the image space.\\

The proposed technique is also compared to other methods in removing the sinogram noise. It is shown that in general, the learning based models return a better structurally correct result in different noise levels compared to model-based techniques. And the proposed method gives better results in background intensities higher than 10000. It is worthwhile to note that the network is not trained on noisy data and adding noise to training samples will increase the robustness of the model to different noise value.\\

The intermediate results have been presented which concludes the fact that the early stages of the model have the most impact on improving the result and also it is shown that both SIRT and DNN cooperate in returning a satisfactory output in both image and sinogram space.\\

As in every other technique, the presented method suffers from several drawbacks explained as follows:\\
\begin{enumerate}
\item The current technique is trained on a specific geometry (20 projections, parallel beam geometry) and will not induce the exact same improvements over other geometries. This issue is not limited to the current method but to every other learning-based technique. In the evaluation section, all other learning-based techniques are trained on the same geometry to accomplish a fair comparison.\\
 
\item The presented technique is slower than other techniques in the evaluation section. This is expectable since the current method deploys the SIRT and DNN blocks several times iteratively. But considering the fast improvement of hardware design, affordability and accessibility of parallel processing machines such as GPUs, this issue is not prohibitive in placing these type of models into the consumer market even with current technologies.\\
\end{enumerate}

The future works include adding sinogram noise to the training data in order to train a more robust model.

\section*{Acknowledgment}
This work is financially supported by VLAIO (Flemish Agency for Innovation and Entrepreneurship), through the ANNTOM project HBC.2017.0595.\\

We gratefully acknowledge the support of NVIDIA Corporation with the donation of a Titan Xp GPU used for this research.

\ifCLASSOPTIONcaptionsoff
  \newpage
\fi



\bibliographystyle{IEEEtran}
\bibliography{bib}

\begin{thebibliography}{10}
\providecommand{\url}[1]{#1}
\csname url@samestyle\endcsname
\providecommand{\newblock}{\relax}
\providecommand{\bibinfo}[2]{#2}
\providecommand{\BIBentrySTDinterwordspacing}{\spaceskip=0pt\relax}
\providecommand{\BIBentryALTinterwordstretchfactor}{4}
\providecommand{\BIBentryALTinterwordspacing}{\spaceskip=\fontdimen2\font plus
\BIBentryALTinterwordstretchfactor\fontdimen3\font minus
  \fontdimen4\font\relax}
\providecommand{\BIBforeignlanguage}[2]{{%
\expandafter\ifx\csname l@#1\endcsname\relax
\typeout{** WARNING: IEEEtran.bst: No hyphenation pattern has been}%
\typeout{** loaded for the language `#1'. Using the pattern for}%
\typeout{** the default language instead.}%
\else
\language=\csname l@#1\endcsname
\fi
#2}}
\providecommand{\BIBdecl}{\relax}
\BIBdecl

\bibitem{mccollough2006ct}
C.~H. McCollough, M.~R. Bruesewitz, and J.~M. Kofler~Jr, ``{CT} dose reduction
  and dose management tools: overview of available options,''
  \emph{Radiographics}, vol.~26, no.~2, pp. 503--512, 2006.

\bibitem{Sidky08}
E.~Y. Sidky and X.~Pan, ``{Image reconstruction in circular cone-beam computed
  tomography by constrained, total-variation minimization},'' \emph{Physics in
  Medicine and Biology}, vol.~{53}, no.~{17}, pp. {4777--4807}, {SEP 7} {2008}.

\bibitem{Donoho06}
D.~Donoho, ``{Compressed sensing},'' \emph{IEEE Transactions on Information
  Theory}, vol.~{52}, no.~{4}, pp. {1289--1306}, {APR} {2006}.

\bibitem{Batenburg11}
K.~J. Batenburg and J.~Sijbers, ``{DART}: A practical reconstruction algorithm
  for discrete tomography,'' \emph{IEEE Transactions on Image Processing}.

\bibitem{Rantala06}
M.~Rantala, S.~Vanska, S.~Jarvenpaa, M.~Kalke, M.~Lassas, J.~Moberg, and
  S.~Siltanen, ``{Wavelet-based reconstruction for limited-angle X-ray
  tomography},'' \emph{IEEE Transactions on Medical Imaging}, vol.~{25},
  no.~{2}, pp. {210--217}, {FEB} {2006}.

\bibitem{he2015delving}
K.~He, X.~Zhang, S.~Ren, and J.~Sun, ``Delving deep into rectifiers: Surpassing
  human-level performance on imagenet classification,'' in \emph{Proceedings of
  the IEEE international conference on computer vision}, 2015, pp. 1026--1034.

\bibitem{ioffe2015batch}
S.~Ioffe and C.~Szegedy, ``Batch normalization: Accelerating deep network
  training by reducing internal covariate shift,'' \emph{arXiv preprint
  arXiv:1502.03167}, 2015.

\bibitem{he2016deep}
K.~He, X.~Zhang, S.~Ren, and J.~Sun, ``Deep residual learning for image
  recognition,'' in \emph{Proceedings of the IEEE conference on computer vision
  and pattern recognition}, 2016, pp. 770--778.

\bibitem{pelt2013fast}
D.~M. Pelt and K.~J. Batenburg, ``Fast tomographic reconstruction from limited
  data using artificial neural networks,'' \emph{IEEE Transactions on Image
  Processing}, vol.~22, no.~12, pp. 5238--5251, 2013.

\bibitem{wu2017iterative}
D.~Wu, K.~Kim, G.~El~Fakhri, and Q.~Li, ``Iterative low-dose {CT}
  reconstruction with priors trained by artificial neural network,'' \emph{IEEE
  transactions on medical imaging}, vol.~36, no.~12, pp. 2479--2486, 2017.

\bibitem{makhzani2015winner}
A.~Makhzani and B.~J. Frey, ``Winner-take-all autoencoders,'' in \emph{Advances
  in Neural Information Processing Systems}, 2015, pp. 2791--2799.

\bibitem{zhu2018image}
B.~Zhu, J.~Z. Liu, S.~F. Cauley, B.~R. Rosen, and M.~S. Rosen, ``Image
  reconstruction by domain-transform manifold learning,'' \emph{Nature}, vol.
  555, no. 7697, p. 487, 2018.

\bibitem{kang2018deep}
E.~Kang, W.~Chang, J.~Yoo, and J.~C. Ye, ``Deep convolutional framelet denosing
  for low-dose {CT} via wavelet residual network,'' \emph{IEEE Transactions on
  Medical Imaging}, vol.~37, no.~6, pp. 1358--1369, 2018.

\bibitem{kang2017wavelet}
E.~Kang, J.~C. Ye \emph{et~al.}, ``Wavelet domain residual network
  ({WavResNet}) for low-dose {X-ray CT} reconstruction,'' \emph{arXiv preprint
  arXiv:1703.01383}, 2017.

\bibitem{chen2017low}
H.~Chen, Y.~Zhang, M.~K. Kalra, F.~Lin, Y.~Chen, P.~Liao, J.~Zhou, and G.~Wang,
  ``Low-dose {CT} with a residual encoder-decoder convolutional neural
  network,'' \emph{IEEE transactions on medical imaging}, vol.~36, no.~12, pp.
  2524--2535, 2017.

\bibitem{chen2017low2}
H.~Chen, Y.~Zhang, W.~Zhang, P.~Liao, K.~Li, J.~Zhou, and G.~Wang, ``Low-dose
  {CT} via convolutional neural network,'' \emph{Biomedical optics express},
  vol.~8, no.~2, pp. 679--694, 2017.

\bibitem{pelt2018improving}
D.~Pelt, K.~Batenburg, and J.~Sethian, ``Improving tomographic reconstruction
  from limited data using mixed-scale dense convolutional neural networks,''
  \emph{Journal of Imaging}, vol.~4, no.~11, p. 128, 2018.

\bibitem{Gregor2008}
J.~Gregor and T.~Benson, ``Computational analysis and improvement of {SIRT},''
  vol.~27, no.~7, pp. 918--924, 2008.

\bibitem{yang2018low}
Q.~Yang, P.~Yan, Y.~Zhang, H.~Yu, Y.~Shi, X.~Mou, M.~K. Kalra, Y.~Zhang,
  L.~Sun, and G.~Wang, ``Low dose {CT} image denoising using a generative
  adversarial network with wasserstein distance and perceptual loss,''
  \emph{IEEE transactions on medical imaging}, 2018.

\bibitem{goodfellow2014generative}
I.~Goodfellow, J.~Pouget-Abadie, M.~Mirza, B.~Xu, D.~Warde-Farley, S.~Ozair,
  A.~Courville, and Y.~Bengio, ``Generative adversarial nets,'' in
  \emph{Advances in neural information processing systems}, 2014, pp.
  2672--2680.

\bibitem{simonyan2014very}
K.~Simonyan and A.~Zisserman, ``Very deep convolutional networks for
  large-scale image recognition,'' \emph{arXiv preprint arXiv:1409.1556}, 2014.

\bibitem{dropout}
N.~Srivastava, G.~Hinton, A.~Krizhevsky, I.~Sutskever, and R.~Salakhutdinov,
  ``Dropout: a simple way to prevent neural networks from overfitting,''
  \emph{The Journal of Machine Learning Research}, vol.~15, no.~1, pp.
  1929--1958, 2014.

\bibitem{dilation}
F.~Yu and V.~Koltun, ``Multi-scale context aggregation by dilated
  convolutions,'' \emph{arXiv preprint arXiv:1511.07122}, 2015.

\bibitem{MSDorig}
D.~M. Pelt and J.~A. Sethian, ``A mixed-scale dense convolutional neural
  network for image analysis,'' \emph{Proceedings of the National Academy of
  Sciences}, vol. 115, no.~2, pp. 254--259, 2018.

\bibitem{relu}
R.~H. Hahnloser, R.~Sarpeshkar, M.~A. Mahowald, R.~J. Douglas, and H.~S. Seung,
  ``Digital selection and analogue amplification coexist in a cortex-inspired
  silicon circuit,'' \emph{Nature}, vol. 405, no. 6789, p. 947, 2000.

\bibitem{astra3}
W.~Palenstijn, K.~Batenburg, and J.~Sijbers, ``Performance improvements for
  iterative electron tomography reconstruction using graphics processing units
  ({GPUs}),'' \emph{Journal of structural biology}, vol. 176, no.~2, pp.
  250--253, 2011.

\bibitem{astra1}
W.~van Aarle, W.~J. Palenstijn, J.~Cant, E.~Janssens, F.~Bleichrodt,
  A.~Dabravolski, J.~De~Beenhouwer, K.~J. Batenburg, and J.~Sijbers, ``Fast and
  flexible x-ray tomography using the {ASTRA} toolbox,'' \emph{Optics express},
  vol.~24, no.~22, pp. 25\,129--25\,147, 2016.

\bibitem{astra2}
W.~van Aarle, W.~J. Palenstijn, J.~De~Beenhouwer, T.~Altantzis, S.~Bals, K.~J.
  Batenburg, and J.~Sijbers, ``The {ASTRA} toolbox: A platform for advanced
  algorithm development in electron tomography,'' \emph{Ultramicroscopy}, vol.
  157, pp. 35--47, 2015.

\bibitem{adam}
D.~P. Kingma and J.~Ba, ``Adam: A method for stochastic optimization,''
  \emph{arXiv preprint arXiv:1412.6980}, 2014.

\bibitem{mxnet}
T.~Chen, M.~Li, Y.~Li, M.~Lin, N.~Wang, M.~Wang, T.~Xiao, B.~Xu, C.~Zhang, and
  Z.~Zhang, ``Mxnet: A flexible and efficient machine learning library for
  heterogeneous distributed systems,'' \emph{arXiv preprint arXiv:1512.01274},
  2015.

\bibitem{teslav100}
\BIBentryALTinterwordspacing
N.~TESLA, ``V100 gpu accelerator,'' \emph{NVIDIA, Oct}, 2016. [Online].
  Available:
  \url{https://images.nvidia.com/content/technologies/volta/pdf/tesla-volta-v100-datasheet-letter-fnl-web.pdf}
\BIBentrySTDinterwordspacing

\bibitem{dgxstation}
\BIBentryALTinterwordspacing
NVIDIA, ``Nvidia dgx station: Ai workstation for data science teams,'' 2018.
  [Online]. Available:
  \url{https://www.nvidia.com/en-us/data-center/dgx-station/}
\BIBentrySTDinterwordspacing

\bibitem{Paige82}
C.~Paige and M.~Saunders, ``{LSQR} - an algorithm for sparse linear-equations
  and sparse least-squares,'' \emph{ACM Transactions on Mathematical Software},
  vol.~8, no.~1, pp. 43--71, 1982.

\bibitem{Yokota17}
T.~Yokota and H.~Hontani, ``{An Efficient Method for Adapting Step-size
  Parameters of Primal-dual Hybrid Gradient Method in Application to Total
  Variation Regularization},'' in \emph{2017 Asia-Pacific Signal and
  Information Processing Association Annual Summit and Conference (APSIPA ASC
  2017)}, ser. {Asia-Pacific Signal and Information Processing Association
  Annual Summit and Conference}, {2017}, pp. {973--979}, {9th Annual Summit and
  Conference of the Asia-Pacific-Signal-and-Information-Processing-Association
  (APSIPA ASC), Kuala Lumpur, MALAYSIA, DEC 12-15, 2017}.

\bibitem{SSIM}
Z.~Wang, A.~C. Bovik, H.~R. Sheikh, E.~P. Simoncelli \emph{et~al.}, ``Image
  quality assessment: from error visibility to structural similarity,''
  \emph{IEEE transactions on image processing}, vol.~13, no.~4, pp. 600--612,
  2004.

\end{thebibliography}
%



%

\begin{IEEEbiographynophoto}{Shabab Bazrafkan}
	received his B.Sc degree from Urmia University, Urmia, Iran in electrical engineering in 2011, M.Sc degree from Shiraz University of Technology (SuTECH) in telecommunication engineering, Image processing branch in 2013 and Ph.D. from the National University of Ireland, Galway (NUIG) in Deep Learning and Neural Network design in 2018 and he is currently a postdoctoral researcher working on low dose CT image reconstruction using machine learning techniques with VisionLab at the University of Antwerp.
\end{IEEEbiographynophoto}
\vskip -2.6\baselineskip plus -1fil
\begin{IEEEbiographynophoto}{Vincent Van Nieuwenhove}
	received his master’s degree in Physics in 2013 at the University of Antwerp, Belgium with a thesis on statistical processing of functional MRI data. Afterwards, he pursued his PhD at the imec-Vision lab, University of Antwerp. He received his PhD in Physics in 2017 with a thesis entitled ‘Model-based reconstruction algorithms for dynamic X-ray CT’. In 2018, Vincent joined Agfa NV, Belgium as ‘Research Engineer in 2-3D Medical Imaging’.
	
\end{IEEEbiographynophoto}
\vskip -2.6\baselineskip plus -1fil
\begin{IEEEbiographynophoto}{Joris Soons}
	received his M.Sc degree in physics from University of Antwerp, Belgium in 2007. During his PhD (2007-2012) at the lab of biomedical physics (University of Antwerp) and as postdoctoral researcher (2012-2015) at the Otobiomechanics group (Stanford University, USA), he focussed on 3D imaging techniques, modelling and mechanical experiments in biomechanics. Currently he is a researcher in 3D reconstruction and image processing at AGFA NV, Belgium.
\end{IEEEbiographynophoto}
\vskip -2.6\baselineskip plus -1fil
\begin{IEEEbiographynophoto}{Jan De Beenhouwer}
	obtained a M.Sc. in Computer Science Engineering in 2003 from the KU Leuven,
	Belgium and a Ph.D. in Biomedical Engineering from the University of Ghent, Belgium in 2008. He was a postdoctoral fellow for 2 years at the same institution prior to joining the Vision Lab at the University of Antwerp, Belgium. Currently, he is a research professor
	and leads the ASTRA group in imec-Vision Lab which focuses on the development of advanced
	computational methods for tomography as well as new reconstruction techniques that lead to better
	reconstruction quality compared to classical reconstruction methods. His main interest is in image
	reconstruction, processing and analysis with focus on computed tomography and electron
	tomography.
\end{IEEEbiographynophoto}
\vskip -2.6\baselineskip plus -1fil
\begin{IEEEbiographynophoto}{Jan Sijbers}
	graduated in Physics in 1993. In 1998, he received a PhD in Physics from the University of Antwerp, Belgium, entitled Signal and Noise Estimation from Magnetic Resonance Images". He was an FWO Postdoc at the University of Antwerp (Belgium) and the Delft University of Technology (the Netherlands) from 2002-2008. In 2010, he was appointed as a senior lecturer at the University of Antwerp, Belgium. In 2014, he became a full professor. He is the head of imec-Vision Lab, a research lab focusing on image reconstruction, processing, and analysis. His main interests are in the domain of Magnetic Resonance Imaging and X-ray Computed Tomography. He is Senior Area Editor of IEEE Transactions on Image Processing as well as Associated Editor of IEEE Transactions on Medical Imaging.
\end{IEEEbiographynophoto}

\end{document}